\documentclass[useAMS,usenatbib]{mn2e}
\usepackage{graphicx} 
\usepackage[usenames,dvipsnames,svgnames,table]{xcolor}
\usepackage{amsmath}
\usepackage{amsfonts}
\usepackage{amssymb}
\usepackage{ulem}

\newcommand{\guy}[1]{{#1}}
\newcommand{\todo}[1]{{#1}}

\newcommand{\Kepler}{\textit{Kepler} }

\title[Asteroseismic inference on rotation in 16 Cyg]{Asteroseismic inference on rotation, gyrochronology and planetary system dynamics of 16 Cygni}
\author[G. R. Davies et al.]{G.R.~Davies$^{1,2,3}$, W.J.~Chaplin$^{2,3}$, W.M.~Farr$^{2}$, R.A.~Garc\'\i a$^{1}$, M.N.~Lund$^{3}$, \newauthor S.~Mathis$^{1}$, T.S.~Metcalfe$^{4,3}$, T.~Appourchaux$^{5}$, S.~Basu$^{6}$, O.~Benomar$^{7}$, \newauthor T.L.~Campante$^{2}$, T.~Ceillier$^{1}$, Y.~Elsworth$^{2}$, R.~Handberg$^{2,3}$, D.~Salabert$^{1}$, D.~Stello$^{8,3}$ \\ 
$^{1}$Laboratoire AIM Paris-Saclay, CEA/DSM -- CNRS - Univ. Paris Diderot -- IRFU/SAp, \\ Centre de Saclay, 91191 Gif-sur-Yvette Cedex, France\\
$^{2}$School of Physics and Astronomy, University of Birmingham, Birmingham, B15 2TT, United Kingdom.\\
$^{3}$Stellar Astrophysics Centre (SAC), Department of Physics and Astronomy, \\ Aarhus University, Ny Munkegade 120, DK-8000 Aarhus C, Denmark\\
$^{4}$Space Science Institute, 4750 Walnut Street, Suite 205, Boulder, CO 80301, USA\\
$^{5}$Institut d'Astrophysique Spatiale, Universit\'{e} Paris 11, CNRS (UMR8617), Batiment 121, F-91405 Orsay Cedex, France\\
$^{6}$Department of Astronomy, Yale University, PO Box 208101, New Haven, CT
06520-8101, USA\\
$^{7}$The University of Tokyo, Tokyo 113-0033, Japan\\
$^{8}$Sydney Institute for Astronomy (SIfA), School of Physics, University of Sydney, NSW 2006, Australia
}
\begin{document}
\date{Accepted 1988 December 15. Received 1988 December 14; in original form 1988 October 11}
\pagerange{\pageref{firstpage}--\pageref{lastpage}} \pubyear{2013}
\maketitle
\label{firstpage}
\begin{abstract}
\guy{The solar analogs 16 Cyg A and 16 Cyg B are excellent asteroseismic targets in the \Kepler field of view and together with a red dwarf and a Jovian planet form an interesting system.  For these more evolved Sun-like stars we cannot detect surface rotation with the current \Kepler data but instead use the technique of asteroseimology to determine rotational properties of both 16 Cyg A and B.  We find the rotation periods to be $23.8^{+1.5}_{-1.8} \rm \, days$ and $23.2^{+11.5}_{-3.2} \rm \, days$, and the angles of inclination to be $56^{+6}_{-5} \, ^{\circ}$ and $36^{+17}_{-7} \, ^{\circ}$, for A and B respectively.  Together with these results we use the published mass and age to suggest that, under the assumption of a solar-like rotation profile, 16 Cyg A could be used when calibrating gyrochronology relations. In addition, we discuss the known 16 Cyg B star-planet eccentricity and measured low obliquity which is consistent with Kozai cycling and tidal theory.}
\end{abstract}
\begin{keywords}
stars: oscillations, stars: rotation, planet-star interactions
\end{keywords}
\section{Introduction}
16 Cyg is a hierarchical triple star system composed of two Sun-like stars in a wide orbit: 16 Cyg A (HD 186408, HR 7503, HIP 96895) and 16 Cyg B (HD 186427,  HR 7504, HIP 96901) together with a red dwarf orbiting component A (16 Cyg C) and a Jovian planet orbiting component B (16 Cyg Bb).  16 Cyg is a well studied system with an extensive literature \citep{1997ApJ...483..457C,1997Natur.386..254H,1999PASP..111..321H}.  Recently, the \Kepler space telescope has observed 16 Cyg A and B and \cite{2012ApJ...748L..10M} have determined accurate fundamental stellar properties using the technique of asteroseismology.\\
Despite extensive observation the stellar rotation of 16 Cyg A and B is not well constrained.  The projected rotation rate ($v \sin i$) cannot be accurately determined from spectroscopic observation\todo{, a result consistent with the expected modest rates of rotation}.  In addition, 16 Cyg A and B are evolved main-sequence stars, with low surface magnetism and a corresponding lack of star spots, complicating measurements of surface rotation.  Estimates of rotational periods have been derived from Ca II H\&K measurements \citep{1991ApJ...375..722S} but these values rely on scaling from other stellar parameters and cannot be considered definitive.\\
A set of well determined rotational parameters for 16 Cyg A and B, in combination with existing measurements, would allow us to test gyrochronology \citep{2007ApJ...669.1167B, 2008ApJ...687.1264M, 2009ApJ...695..679M, 2010ApJ...719..602S} in a region of the HR diagram not normally accessible to classical methods.  In addition, such a measurement constrains the star-planet obliquity in the 16 Cyg Bb system.\\
To study the stellar rotation of 16 Cyg A and B we have examined the \Kepler asteroseismic data sets for the signatures of stellar rotation.  The very high asteroseismic signal-to-noise ratios in both 16 Cyg A and B make this system an excellent candidate to study.  Based on the observed rotational splitting, we estimate \guy{the stellar rotation rate} and the angle of inclination of the rotation \guy{axis} relative to the line of sight.  We use this new information on the period of rotation to test two gyrochronology relations.  In addition, we use the estimated angle of inclination of 16 Cyg B to determine the projected obliquity and hence discuss possible Kozai cycling of the planet-star system.
\section{Adopted properties of the system}
Table 1 shows the properties of the stellar and planetary components adopted for this work.  The asteroseismic results are taken from \cite{2012ApJ...748L..10M}, their results being a collaborative effort using a number of different stellar evolution codes and detailed asteroseismic modelling approaches.  \guy{The adopted age is consistent with other asteroseismic results from \cite{2013MNRAS.435..242G}, $\approx 6.5 \pm 0.3 \rm \, Gyr$, and \cite{2014ApJ...790..138V}, $6.6 \pm 0.3 \rm \, Gyr$.}  Determination of the age of the system from spectroscopic parameters and isochrone fitting has produced age estimates greater than the asteroseismic results (e.g. $9.1 \pm 0.8 \, \rm Gyr$ from \cite{2003AJ....125.2664L} or $8.0 \pm 1.4 \, \rm Gyr$ from \cite{1998A&A...336..942F}) but more recently \cite{2014A&A...562A..92D} give an age for 16 Cyg B of $6.23 \, \rm Gyr$ that is consistent with the seismic result.\\  
\cite{1981A&A....94..207P} give $B-V$ values for the A and B components\guy{, $0.64$ and $0.66$ respectively}.  In the absence of a quoted precision we adopt uncertanties on $B-V$ as $\pm 0.01$ which more than encompasses the spread in results from other observations \citep{1953ApJ...117..361J,1966MNRAS.133..475A, 1979PASP...91..180M}.  Accuracy or precision greater than this level is not required in this study.\\
\begin{table*}
 \label{tab::adopted}
\caption[Caption]{Adopted properties of the system.} 
\begin{tabular}{lc}
  \begin{tabular}{cccc}
  \hline
  & 16 Cyg A & 16 Cyg B & 16 Cyg Bb \\
  \hline \hline
  Age (Gyr) & $6.8 \pm 0.4^{\rm a}$ & $6.8 \pm 0.4 ^{\rm a}$ & $6.8 \pm 0.4$ \\
  Mass  & $1.11 \pm 0.02^{\rm a}$ $\rm M_{\odot}$ & $1.07 \pm 0.02 ^{\rm a}$ $\rm M_{\odot}$ & $2.38 \pm 0.04 ^{\rm c}$ $\rm M_{\rm Jup}$ \\
  Radius  & $1.243 \pm 0.008 ^{\rm a}$ $\rm R_{\odot}$ & $1.127 \pm 0.007 ^{\rm a}$ $\rm R_{\odot}$ & - \\
  $B-V$ & $0.64^{\rm b} \pm 0.01$ & $0.66 ^{\rm b} \pm 0.01$ & n/a \\
  $\rm T_{eff}$ & $5825 \pm 50^{\rm e}$ (G2V) & $5750 \pm 50^{\rm e}$ (G2V) & n/a \\
  $\left[ \rm Fe/H \right]$ & $0.096 \pm 0.026^{\rm e}$ & $0.052 \pm 0.021^{\rm e}$ & n/a \\
  Orbital Period & $>13000^{\rm c} \rm \; yr$ & $>13000^{\rm c} \rm \; yr$ & $798.5 \pm 1.0 ^{\rm c}$ days\\
  Eccentricity & 0.54 to 1$^{\rm d}$ & 0.54 to 1$^{\rm d}$ & $0.689 \pm 0.011 ^{\rm c}$ \\
  Orbital Inclination ($^{\circ}$)& 100 to 160$^{\rm d}$ & 100 to 160$^{\rm d}$ & $45/135 ^{\rm c}$ \\ 
  \hline \\
  \end{tabular} \\
$\rm ^{a}$ \protect\cite{2012ApJ...748L..10M} ; $\rm ^{b}$ \protect\cite{1981AnA....94..207P} ; $\rm ^{c}$ \protect\cite{2013AJ....146..108P} ; \\ $\rm ^{d}$ \protect\cite{1999PASP..111..321H} ; \todo{ $\rm ^{e}$ \protect\cite{2009AAA...508L..17R}}.\\
\end{tabular}
\end{table*}
The planet 16 Cyg Bb is in an eccentric orbit around the star 16 Cyg B and the inclination of the orbit ($i = 45/135 \pm 1 \, ^{\circ} $) has been estimated as part of a three-body problem \citep{2013AJ....146..108P}.  It is not clear whether 16 Cyg A induces the observed eccentricity in the planetary orbit \citep{1997Natur.386..254H} or not \citep{1999PASP..111..321H}. And as only a single planet has been detected, if planet-planet scattering is responsible for the orbit of 16 Cyg Bb, then the other planets in the system must have been ejected through scattering events \citep{2008ApJ...678..498N, 2012ApJ...751..119B}. 
\section{Data preparation}
Both 16 Cyg A and B are brighter (V $\sim$ 6) than the saturation limit for which \Kepler observations were designed. However, it was possible to capture the full stellar flux by using custom photometric aperture masks. Thus, 928 days of short-cadence observations \citep{2010ApJ...713L.160G} - from Quarter 7 to 16 - were generated using simple aperture photometry \citep{2010ApJ...713L..87J} and then corrected for instrumental perturbations following the methods described by \cite{2011MNRAS.414L...6G}. The final light curves used for asteroseismic analyses were high-pass filtered using a triangular smooth of 4 days width and have a duty cycle of 90.5 $\%$. The power density spectra were computed using a Lomb-Scargle algorithm.\\

\section{Asteroseismic determination of rotation}
Asteroseismic estimation of stellar rotation rests on our ability to detect the signature of rotation in the non-radial modes of the frequency-power spectrum.  Excellent constraints can be found for the rotational splitting multiplied by the sine of the angle of inclination, i.e. the projected rotational splitting \citep{2006MNRAS.369.1281B}, and good constraints can be found for the angle of inclination.  Here we followed the procedure set out in \cite{2013ApJ...766..101C} to estimate the asteroseismic rotation.\\
The desired rotation estimates are outputs of `peak bagging' \citep{2003Ap&SS.284..109A}, which is modelling of the observed power spectrum.  We modelled a background with the sum of two Harvey-like components \citep{1985ESASP.235..199H}.  Modes of oscillation are modelled as a sum of Lorentzian profiles that characterise the power limit spectrum of stochastically excited and intrinsically damped modes.  Peak bagging was performed using Markov Chain Monte Carlo (MCMC) methods (see \cite{2009A&A...506...15B} and \cite{2011A&A...527A..56H}).\\  

\section{Results}
\begin{figure}
\includegraphics[width=88mm,clip]{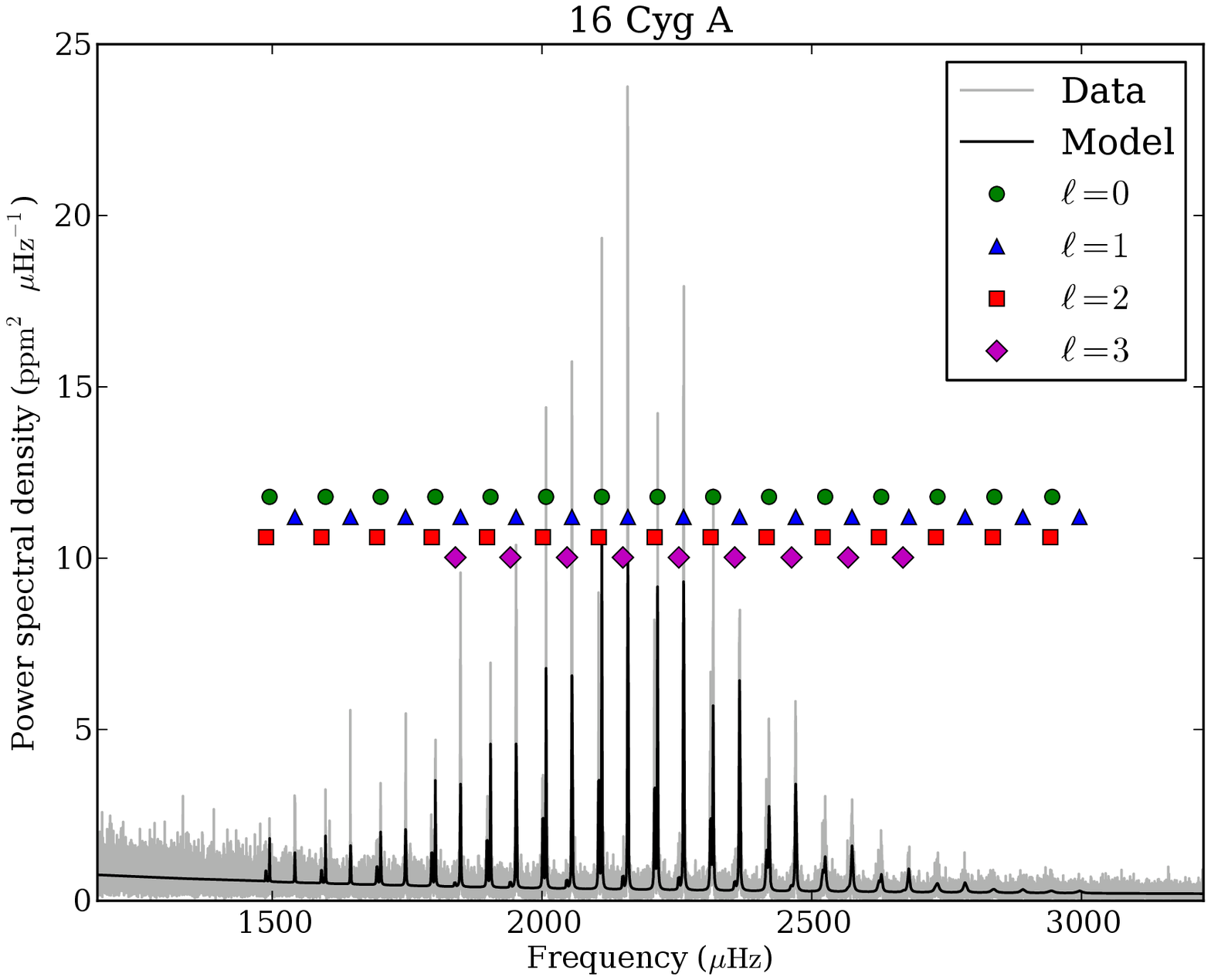}
\includegraphics[width=88mm,clip]{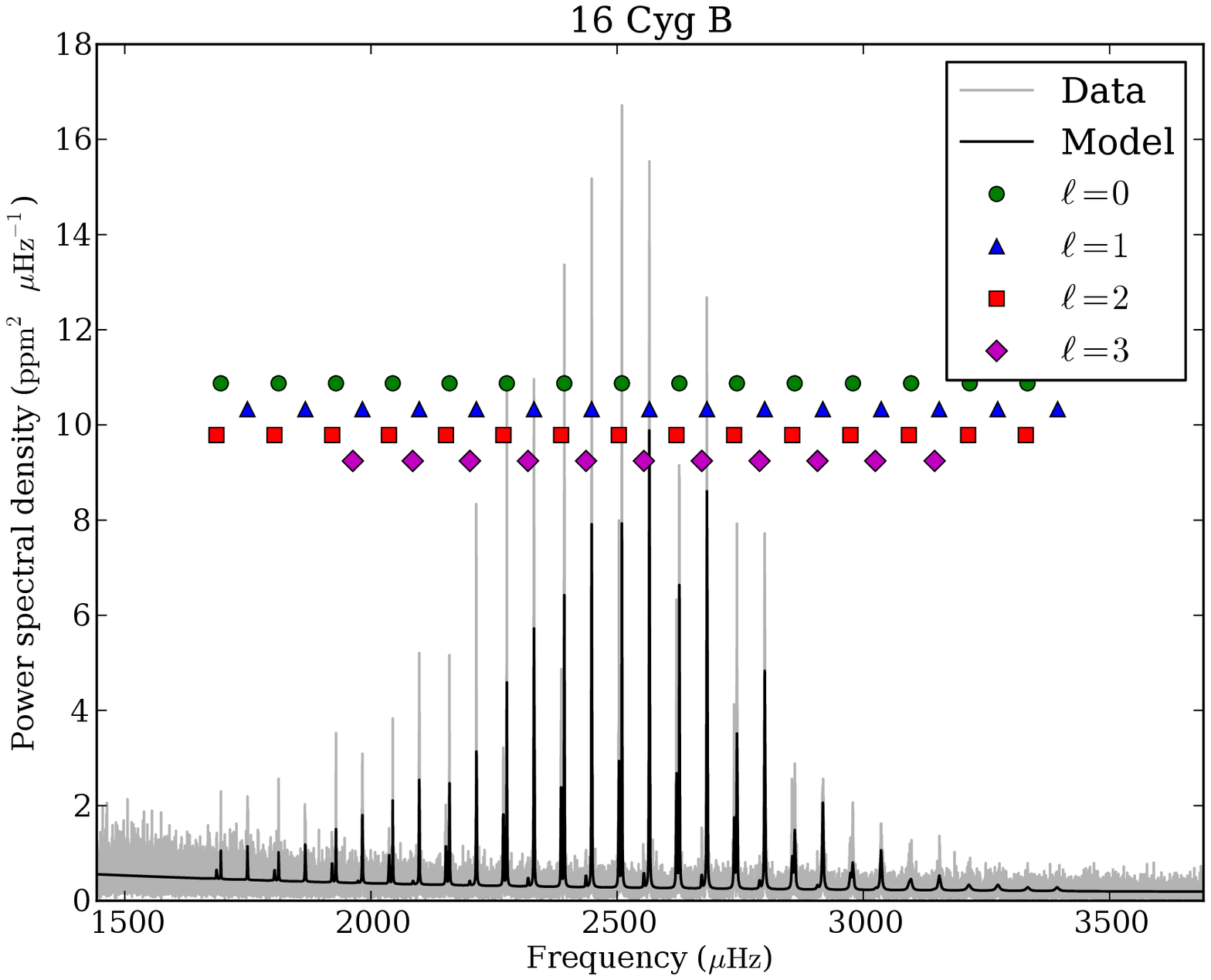}
  \caption{Power spectra in the region of modes of oscillation.  Top: 16 Cyg A; Bottom: 16 Cyg B.  The data are presented along with a fitted model of the power spectrum and labels for each mode of oscillation considered.} 
  \label{fig::psd}
\end{figure}
\begin{figure}
\includegraphics[width=88mm,clip]{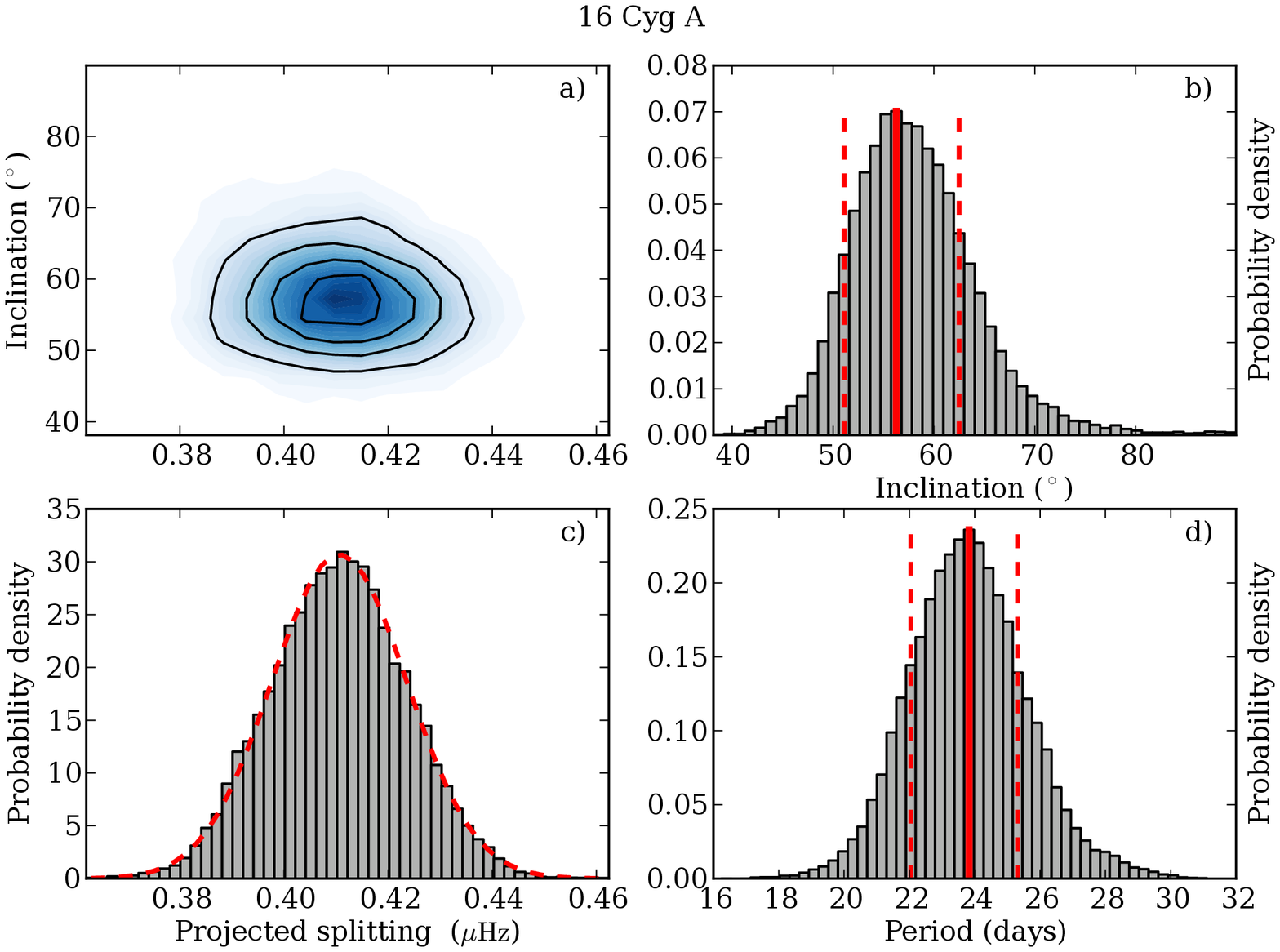}
\includegraphics[width=88mm,clip]{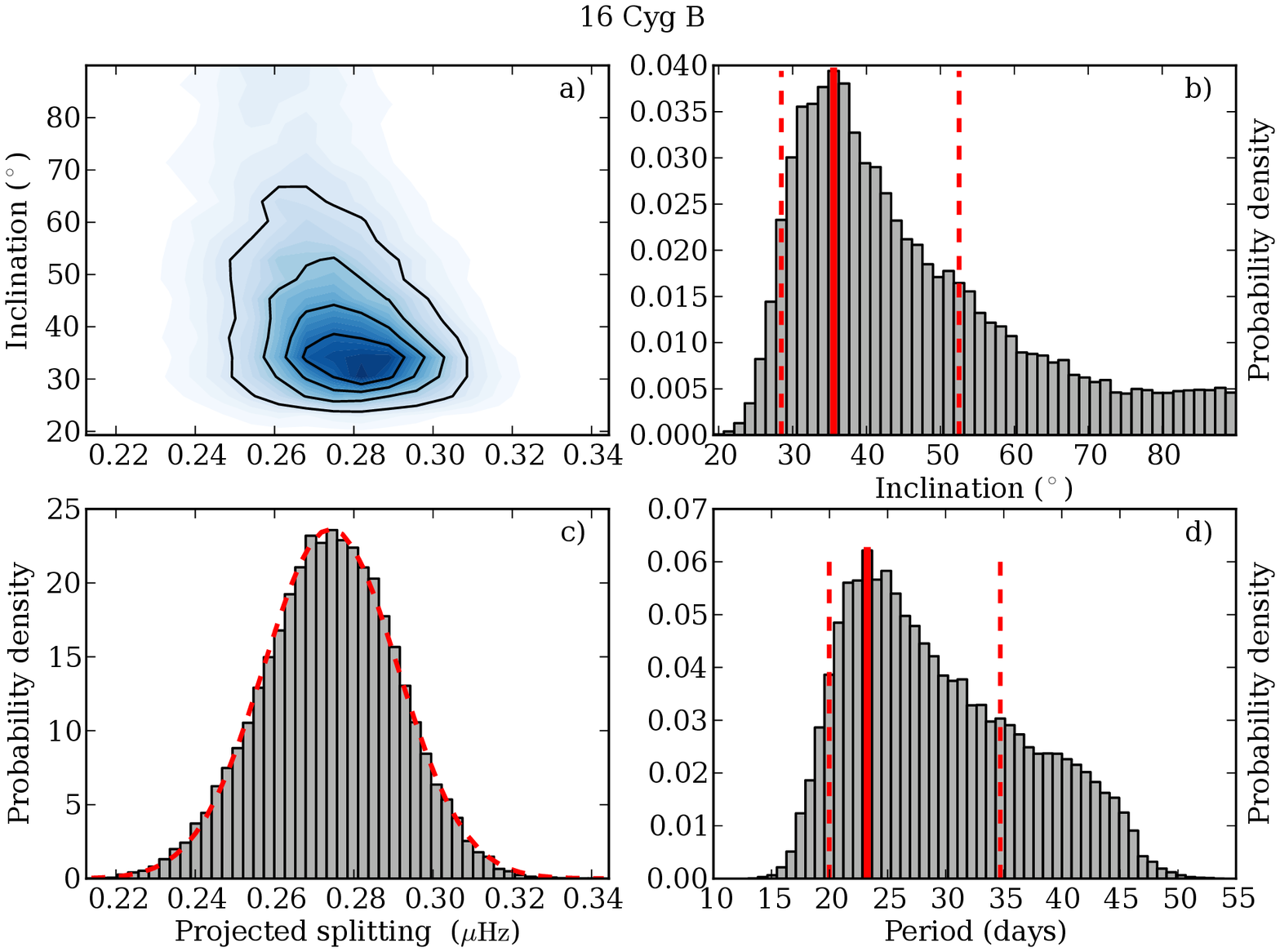}
  \caption{Posterior probability distributions for stellar rotation.  Top: 16 Cyg A; Bottom: 16 Cyg B.  Each quad plot is composed of: a) inclination vs projected splitting PPD; b) projected inclination PPD; c) projected splitting PPD; and d) period of rotation PPD.} 
  \label{fig::rotation}
\end{figure}

\guy{Figure \ref{fig::psd} shows the \Kepler power spectra for 16 Cyg A and B together with the best fitting models of the modes of oscillation and background.  Figure \ref{fig::psd} demonstrates the excellent asteroseismic signal-to-noise ratios for both stars which allows us to extract the signatures of rotation.\\}
The results from our analysis of the internal rotation of both stars are given in Figure \ref{fig::rotation} as the marginalised posterior probability distributions (PPD).  We present both the 2D PPD of projected splitting versus angle of inclination and the corresponding 1D PPD's and the rotational period 1D PPD.\\
The projected splitting is well described by a normal distribution but the angle of inclination is more complicated.  For the projected splitting we report the median value together with the standard deviation which defines the $68\%$ credible region.  For the angle of inclination we report the mode of the distribution together with the $68\%$ highest posterior density (HPD) intervals.  In tests on simulated data we find the mode to be a better estimate of the simulated inclination than other summary statistics (e.g. mean or median).\\
Figure \ref{fig::rotation} also shows the PPD's for rotational period equivalent to the measured rotation.  Again, the PPD's are best described by the modes of the distributions together with the 68\% HPD credible regions.  Table 2 contains the statistical descriptions of the PPD's.\\
\begin{table}
 \begin{minipage}{80mm}
  \caption{Summary statistics for stellar rotation.}
  \begin{tabular}{ccc}
  \hline
  & 16 Cyg A & 16 Cyg B \\
  \hline \hline
  Projected splitting ($\rm \mu Hz$) & $0.411 \pm 0.013 $ & $0.274 \pm 0.017$ \\ 
  Angle of inclination ($^{\circ}$) & $56_{-5}^{+6}$ & $36_{-7}^{+17}$ \\
  Rotational Period (days) & $23.8_{-1.8}^{+1.5}$ & $23.2_{-3.2}^{+11.5}$ \\
  Asteroseismic $v \sin i$ ($\rm km \; s^{-1}$) & $2.23 \pm 0.07$ & $1.35 \pm 0.08$ \\
  \hline \\
 \end{tabular}
\end{minipage}
\label{tab::rotation}
\end{table}
We estimate the asteroseismic $v \sin i$ from the projected splitting and the adopted radius, i.e. $v \sin i = 2 \pi R \, \left< \delta \nu_{s} \right> \sin i$.  Of course, the rotation we estimate asteroseismically is not surface rotation. 
\subsection{Independent validation with the Sun}
\begin{figure}
\includegraphics[width=88mm,clip]{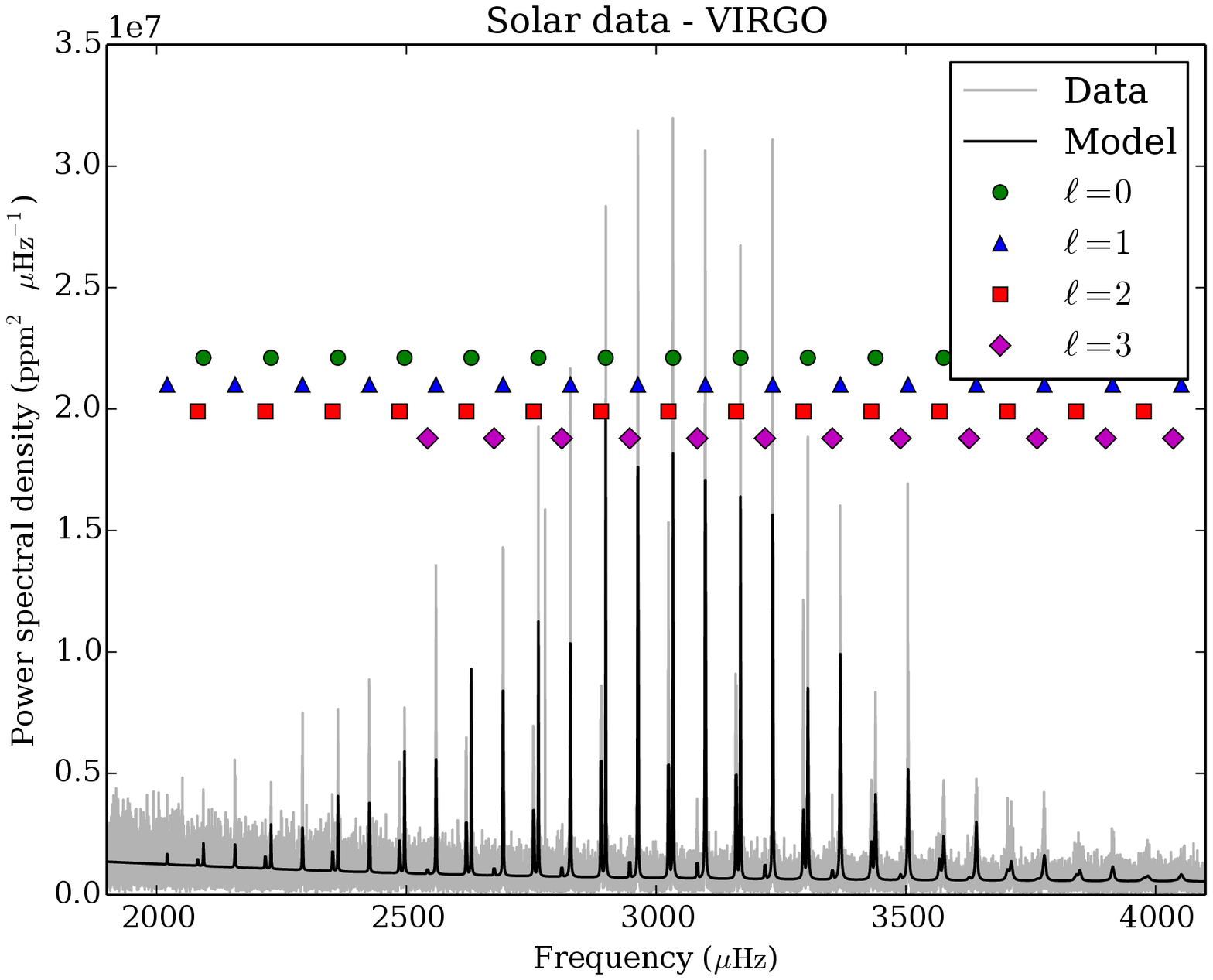}
  \caption{Power spectra in the region of solar modes of oscillation.  The data are presented along with a fitted model of the power spectrum and labels for each mode of oscillation considered.} 
  \label{fig::virgo_psd}
\end{figure}

\begin{figure}
\includegraphics[width=88mm,clip]{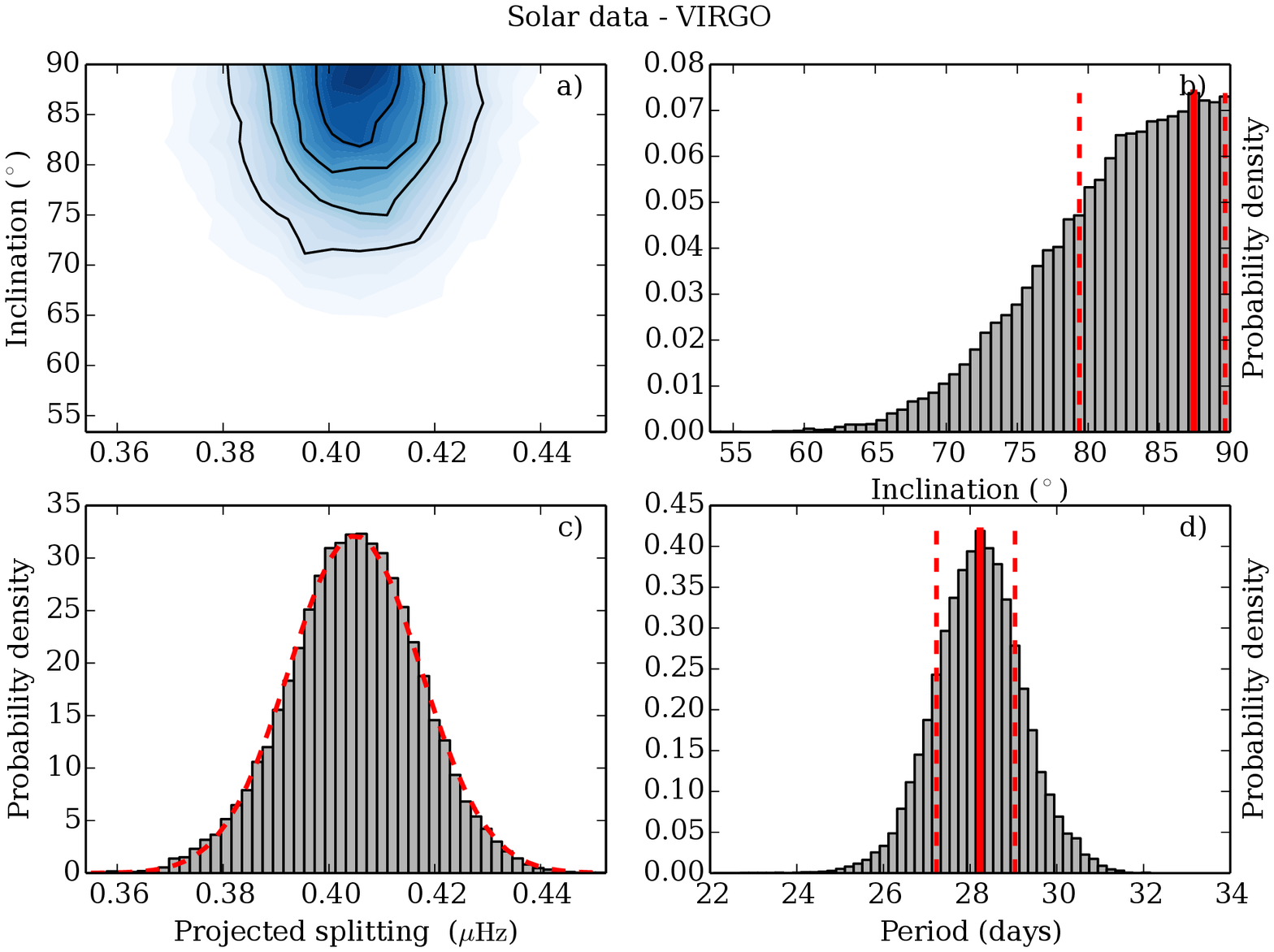}
  \caption{Posterior probability distributions for solar rotation. Composed of: a) inclination vs projected splitting PPD; b) projected inclination PPD; c) projected splitting PPD; and d) period of rotation PPD.} 
  \label{fig::virgo}
\end{figure}

The seismic Sun-as-a-star rotation is well defined and has been accurately measured \citep{2008SoPh..251..119G,2014MNRAS.439.2025D} and hence we can check our asteroseismic method by determining rotation on the Sun.  Using data provided by the red channel of the SPM/VIRGO (triple Sun PhotoMeter/Variability of solar IRradiance and Gravity Oscillations) instrument \citep{1995SoPh..162..101F} on board the SoHO mission \citep{1995SoPh..162....1D}, with white noise added to levels comparable with 16 Cyg, we have performed the above analysis.  Figure \ref{fig::virgo_psd} shows the VIRGO power spectrum with added noise and Figure \ref{fig::virgo} shows the solar rotation PPD's.  The returned period of rotation is $28.2_{-1.0}^{+0.9} \, \rm days$ which is consistent with the Carrington rotation rate (synodic period) of $27.2753 \, \rm days$.\\

\subsection{Surface activity \guy{and rotation}}
In order to identify signals present due to surface rotation we have analysed both stars in the same manner as \cite{2014A&A...562A.124M} and \cite{2014arXiv1403.7155G}.  There are signals across the periodogram and particularly in the region of 20 to 30 days but we cannot determine if they are genuine stellar signals or if they are due to a ``pollution'' related to the \Kepler months. Therefore, we conclude that we cannot detect surface rotation of these stars from the current \Kepler long-cadence light curves.\\
We have analysed data on the Ca II H\&K line from both stars taken at the Lowell observatory from 5 December 1993 to 9 August 2012 (J. Hall, private communication).  Once again, analysis of this data does not produce unambiguous detection of a surface rotation signal.  Tantalisingly, the 16 Cyg A H\&K data show a prominent but not conclusive peak at a period of 26 days but only when considering data collected after 2008, the date of a marked improvement in the quality of the H\&K data.\\
\cite{2013arXiv1311.3374M} give chromospheric activity $S$ indices of $0.1556 \pm 0.0011$ and $0.1537 \pm 0.0005$ for A and B respectively, representing low-levels of activity.  The lack of an unambiguous surface activity signature is consistent with the expected low surface magnetic activity of a late main-sequence star.

\section{Discussion}
\guy{
\subsection{On the comparison of asteroseismic and surface rotation}
Asteroseismic rotation is a measure of internal rotation, that is, a different measure of rotation to the normally used surface rotation measured for gyrochronology.  Here we have validated our asteroseismic rotation method on the Sun and we find that the asteroseismic rotation period estimate is in agreement (${\sim}1\sigma$) with the known solar surface rotation period.  Other studies have found good agreement in Sun-like stars between the asteroseismic rotation and surface rotation measures \citep{2013ApJ...766..101C, 2013PNAS..11013267G}.  The trivial solution that can explain this consistency between different measures of rotation is that there is an absence of significant differential rotation \citep[see][for a general discussion of what is significant]{2014ApJ...790..121L}.  Here we discuss the internal sensitivity of the measured asteroseismic rotation and whether or not it is sensible to use the measured asteroseismic rotation period as a proxy for surface rotation in stars very much like the Sun.\\
The frequency shift of a rotationally split mode component, (labelled by radial order $n$, degree $\ell$, and azimuthal order $m$) is proportional to the integral over the internal rotation profile, $\Omega(r, \theta)$, multiplied by the rotation kernel, $\mathcal{K}_{n\ell m}(r, \theta)$, where $r$ is the radial coordinate, $\theta$ is colatitude, and $R$ is the stellar radius.  The frequency shift is then \citep[e.g., ][]{1977ApJ...217..151H},
\begin{equation}
\delta \omega_{n\ell m} = m \int_{0}^{R} \int_{0}^{\pi} \mathcal{K}_{n\ell m}(r, \theta) \Omega(r, \theta) 
\, r \; {\rm d}r \; {\rm d}\theta.
\label{eqn::2drot}
\end{equation}
It is clear then that for $m > 0$, the measured asteroseismic rotation is determined by both the stellar internal rotation profile and the sensitivity of the mode of oscillation to the stellar interior.\\
\begin{figure*}
\includegraphics[width=0.3\textwidth]{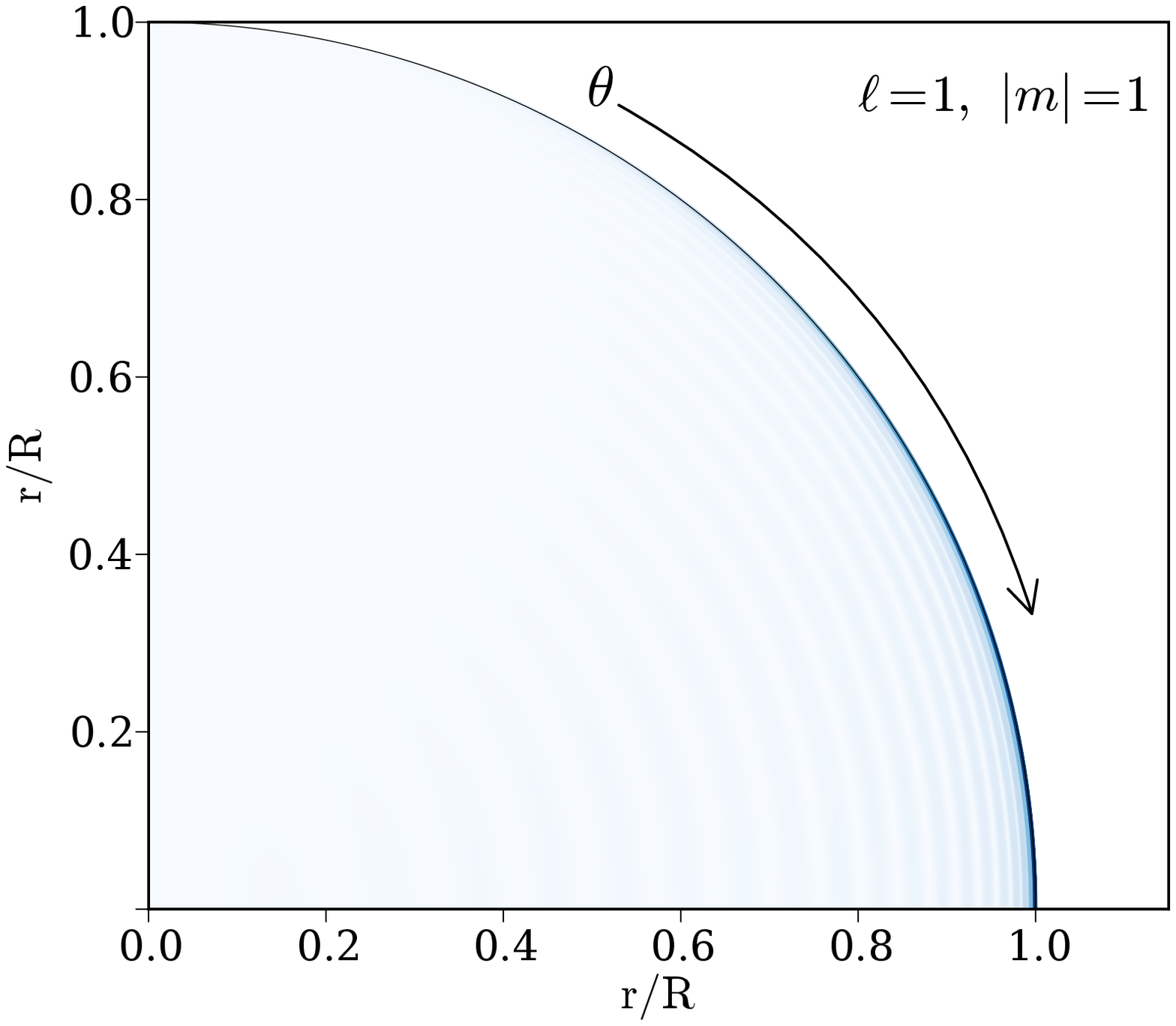}
\includegraphics[width=0.3\textwidth]{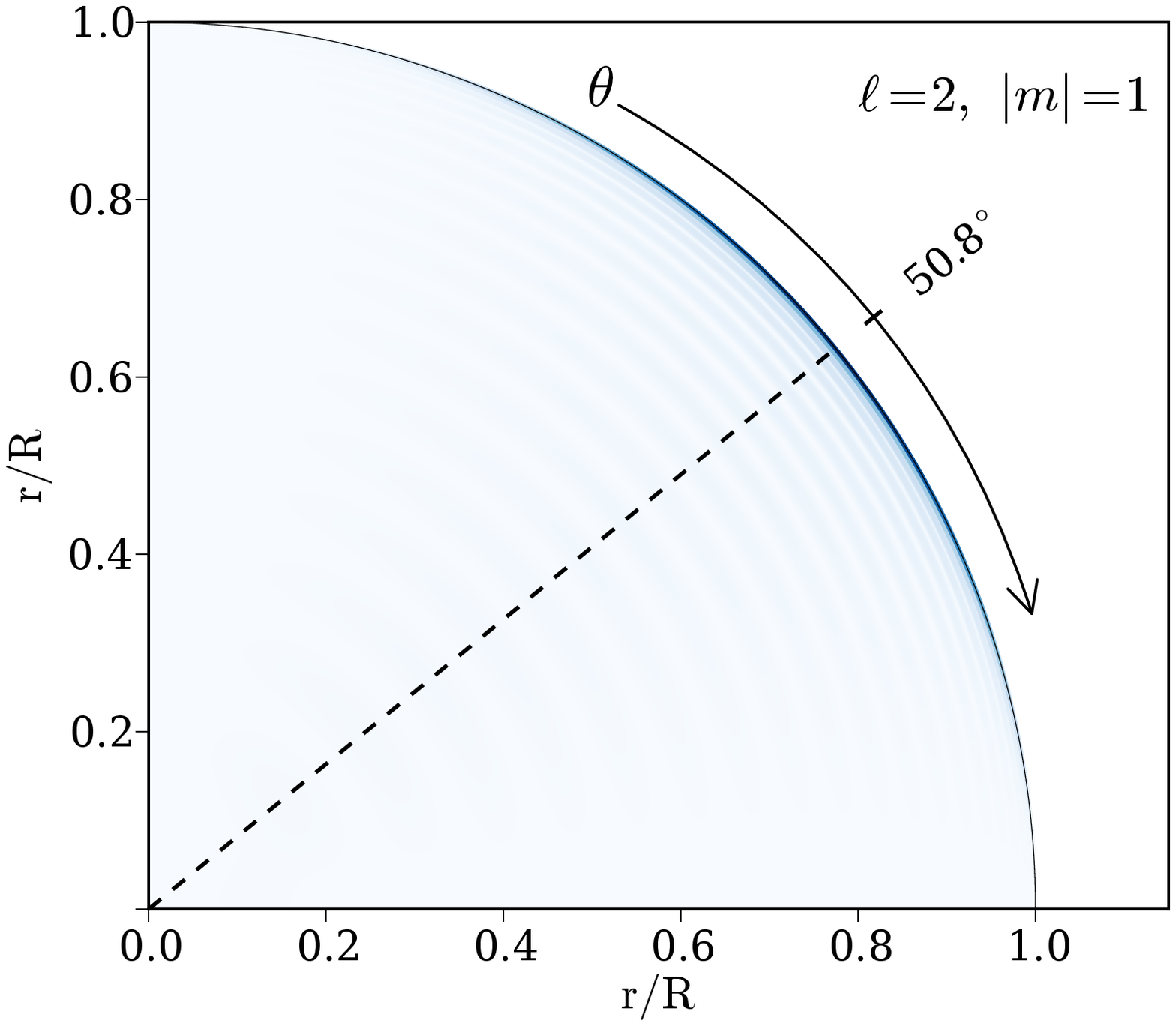}
\includegraphics[width=0.3\textwidth]{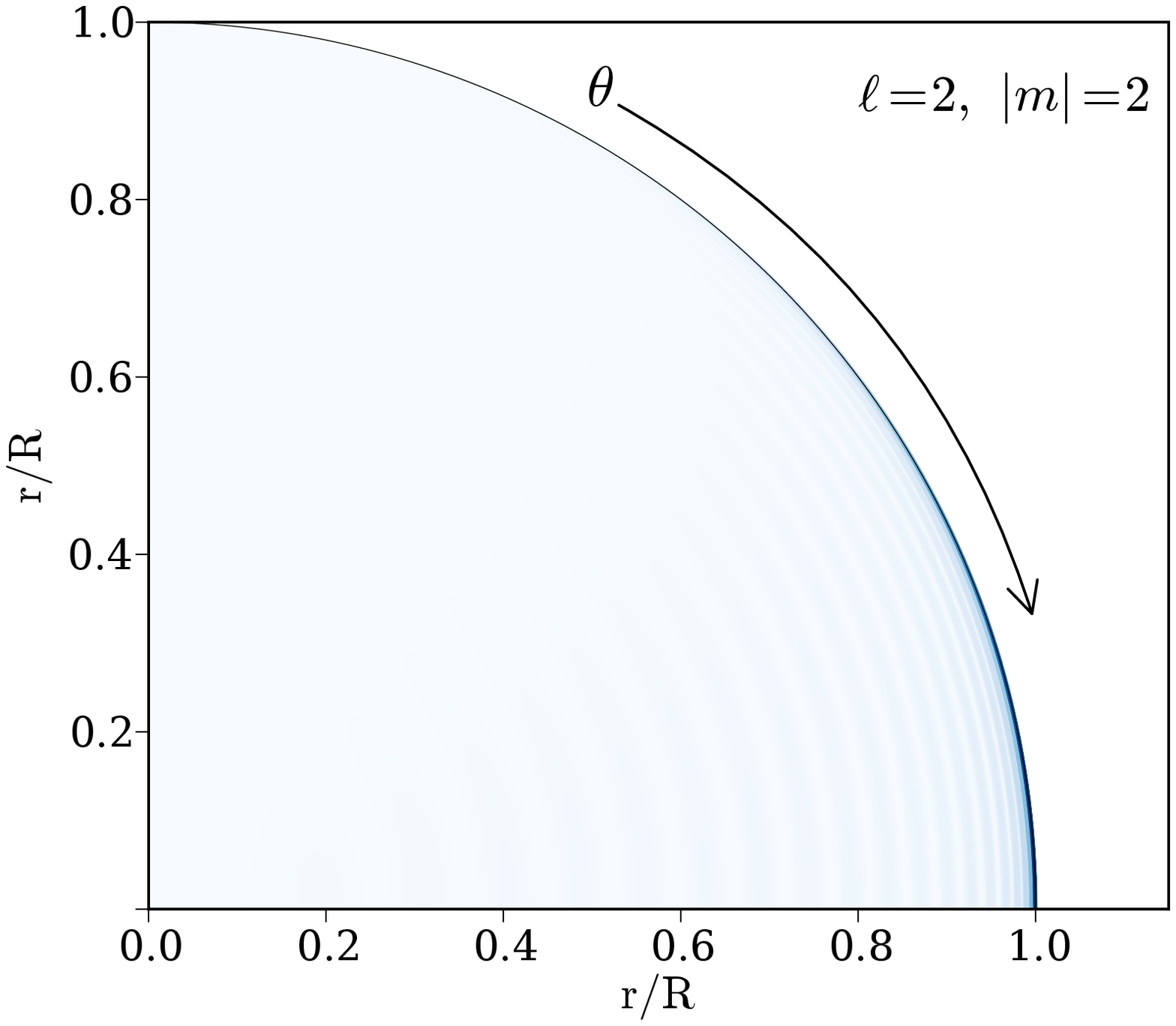}
\caption{Contour plots of rotational kernels for the 16 Cyg A model (see Table \ref{tab::adopted}), with intensity (and thus rotational sensitivity) going from white (low) to blue (high) on a linear colour-scale. Kernels are shown for modes of degree $\ell=1$ and 2, and radial order $n=20$ (corresponding to model frequencies of $\rm {\sim}2267\, \mu Hz$ ($\ell=1$) and $\rm {\sim}2317\, \mu Hz$ ($\ell=2$)). Only one quadrant of the star is shown and in units of the stellar radius. The displayed kernel in each tile may by mirrored in both axes. For the kernel with $\ell=2$ and $|m|=1$ the maximum in co-latitude is different from the equator ($\theta=\,$90$\, ^{\circ}$) and has been indicated by a dashed line.}
\label{fig::2dkernels}
\end{figure*}
Figure \ref{fig::2dkernels} shows the rotation kernels for three different rotationally split components taken from a model \citep{2008Ap&SS.316...13C, 2008Ap&SS.316..113C} of 16 Cyg A.  On the scale shown here, the appearance of kernels for the same $\ell$ and $m$ are largely indistinguishable for the different $n$ measured in this paper.  It is clear that the peak sensitivity to rotation is located very close to the surface but sensitivity to latitude varies with $\ell$ and $|m|$.\\
From Equation \ref{eqn::2drot} there is no rotational splitting for $m = 0$ components.  For components with $|m| = \ell$ the peak sensitivity is located near the surface and at the equator.    For modes with $\ell=2, |m|=1$ the peak of sensitivity is located near the surface and close to mid latitude.  In this work, although we have considered $\ell=3$ modes in our model, the vast majority of constraint is obtained from $\ell=1$ and $2$ modes and so in what follows we will neglect the $\ell=3$ modes.\\
One should note that the kernels, although peaked in sensitivity near the surface and at certain positions in latitude, have some sensitivity down to $r \sim 0.1 R$ and across virtually all latitudes.  It is then possible to conceive of a scenario where differential rotation could manifest as a significant difference between measured surface and asteroseismic rotation periods.  There are at least three possible causes of differences: ``fast'' core rotation, i.e. strong radial differential rotation; ``surface'' latitudinal differential rotation i.e. as observed from star spots; a ``tachocline'' break here defining both a radial and latitudinal differential rotation that is as observed in the Sun.   
\subsubsection{Radial dependence}
Let us consider the radial dependence in isolation.  Rotational splitting with the latitudinal dependence removed is given as \citep[e.g., ][]{1977ApJ...217..151H},
\begin{equation}
\delta \omega_{n\ell m} = m \beta_{n\ell} \int_{0}^{R} \mathcal{K}_{n\ell}(r) \Omega(r) \; {\rm d}r,
\end{equation}
where
\begin{equation}
C_{nl} = 1 - \beta_{n\ell}
\end{equation}
and $-mC_{n\ell}\Omega$ is the rotational splitting resulting from the coriolis force.  For high-order p modes, as we are analysing here, $\beta_{n\ell} \approx 1$ but we have used values calculated from the stellar model.\\ 
\begin{figure}
\includegraphics[width=88mm,clip]{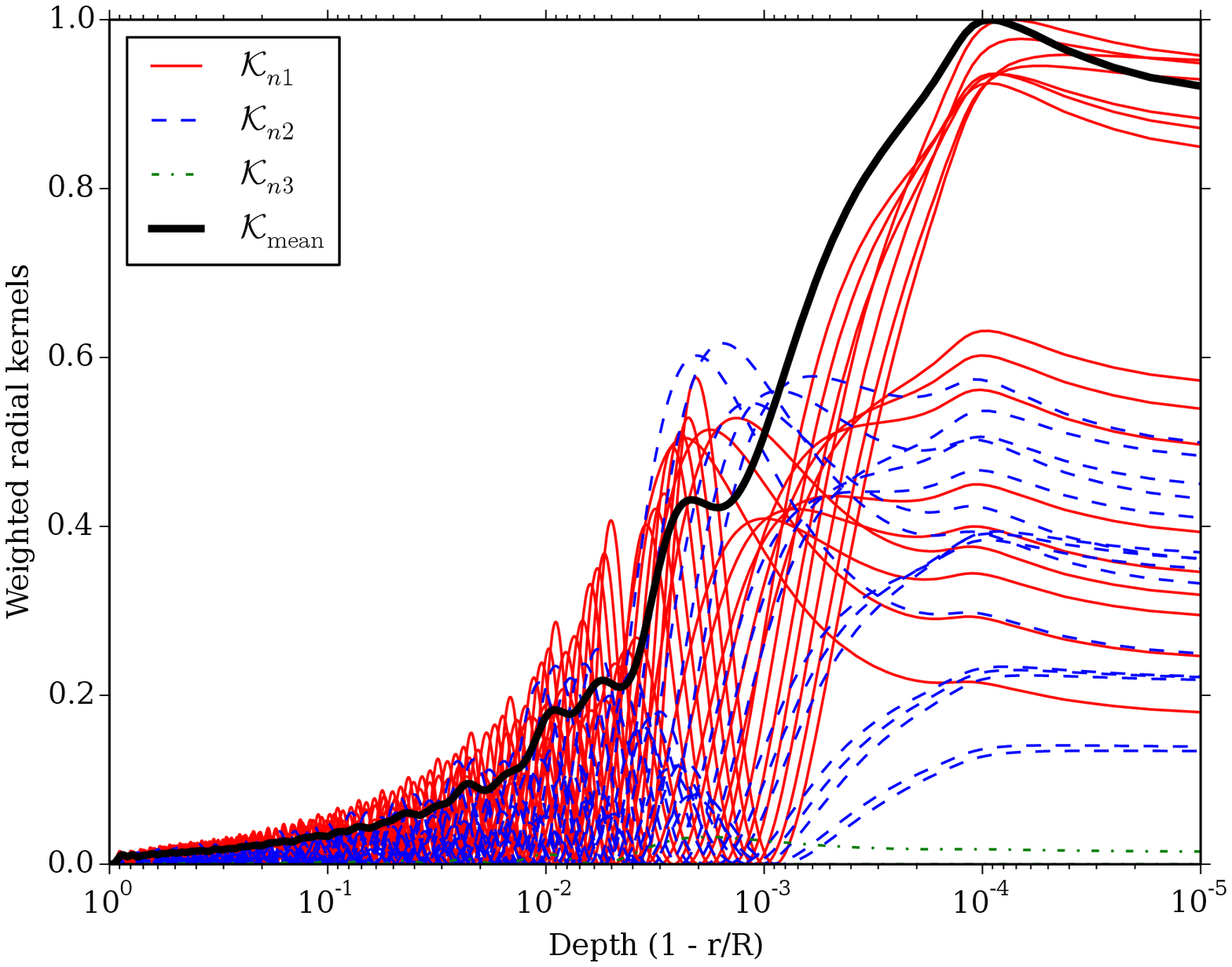}
  \caption{Radial rotation kernels, $\mathcal{K}_{n\ell}$, as a function of depth for 16 Cyg A.  Each kernel is weighted by the reciprocal of the uncertainty in mode frequency and normalised such that the maximum value of all the individual kernels is unity.  The strong black line shows the weighted mean of the individual kernels that represents the radial rotational kernel for the averaged frequency splitting.  Note that we follow the common convention to place the centre of the star on the left hand side and the near surface on the right.} 
  \label{fig::kernels}
\end{figure}
In this work we measure the averaged seismic rotation $\langle \delta \omega_{n\ell} \rangle$.  Each non-radial mode of oscillation provides some contribution to the measured value.  Here we simplify the contribution to the measured rotation of each mode as the reciprocal of the uncertainty estimate of the mode frequency, $\sigma_{n\ell}$, giving the weight as
\begin{equation}
W_{n\ell} = \frac{1}{\sigma_{n\ell}}.
\label{eqn::weight}
\end{equation}
The $68 \%$ credible regions for each fitted mode are listed in Table \ref{tab::freqa}.  This then gives us a proxy for our measure of averaged rotational splitting as the weighted average of all splittings,
\begin{equation}
\langle \delta \omega_{n\ell} \rangle = \frac{\sum_{n} \sum_{\ell} W_{n\ell} \; \delta \omega_{n\ell} }{\sum_{n} \sum_{\ell} W_{n\ell}}.
\end{equation}
Figure \ref{fig::kernels} shows radial mode kernels weighted following equation \ref{eqn::weight} and normalised such that the maximum of all the mode kernels is equal to unity.  On this logarithmic scale it is possible to discern the differences between different $n$ and $\ell$.  It is clear that the majority of constraint on rotation comes from the $\ell=1$ and then the $\ell=2$ modes, while the $\ell=3$ modes have little impact.\\
With the above we have enough information to test rotation profiles in radius and assess the extent to which an estimate of asteroseismic rotation period would differ from that of surface rotation.  Let us consider a rotational profile that has some central rotation angular velocity $\Omega_{\rm c}$ and envelope $\Omega_{\rm e}$.  We define a simple step profile with two levels profile with a discontinuity in rotation rate at $r = \alpha R$ where $0 \leq \alpha \leq 1$ with,
\begin{equation}
\Omega(r) = \begin{cases} 
	\Omega_{\rm c} & \text{if } r < \alpha R \\
	\Omega_{\rm e} & \text{if } r \geq \alpha R.
\end{cases}
\end{equation}
Figure \ref{fig::fastcore} shows the equivalent of the measured averaged asteroseismic rotational splitting for different values of $\alpha$ and $\Omega_{\rm c} / \Omega_{\rm e}$.  The region bounded by the 6\% line represents the parameter space for which the measured asteroseismic rotation would lie within the $1\sigma$ uncertainty for 16 Cyg A.  For all modest rates of core rotation rate, as would be anticipated for a Sun-like star like 16 Cyg A, the bias on the measured rotation is encompassed by the $1 \sigma$ uncertainty.\\
\begin{figure*}
\centering
\includegraphics[width=88mm, clip]{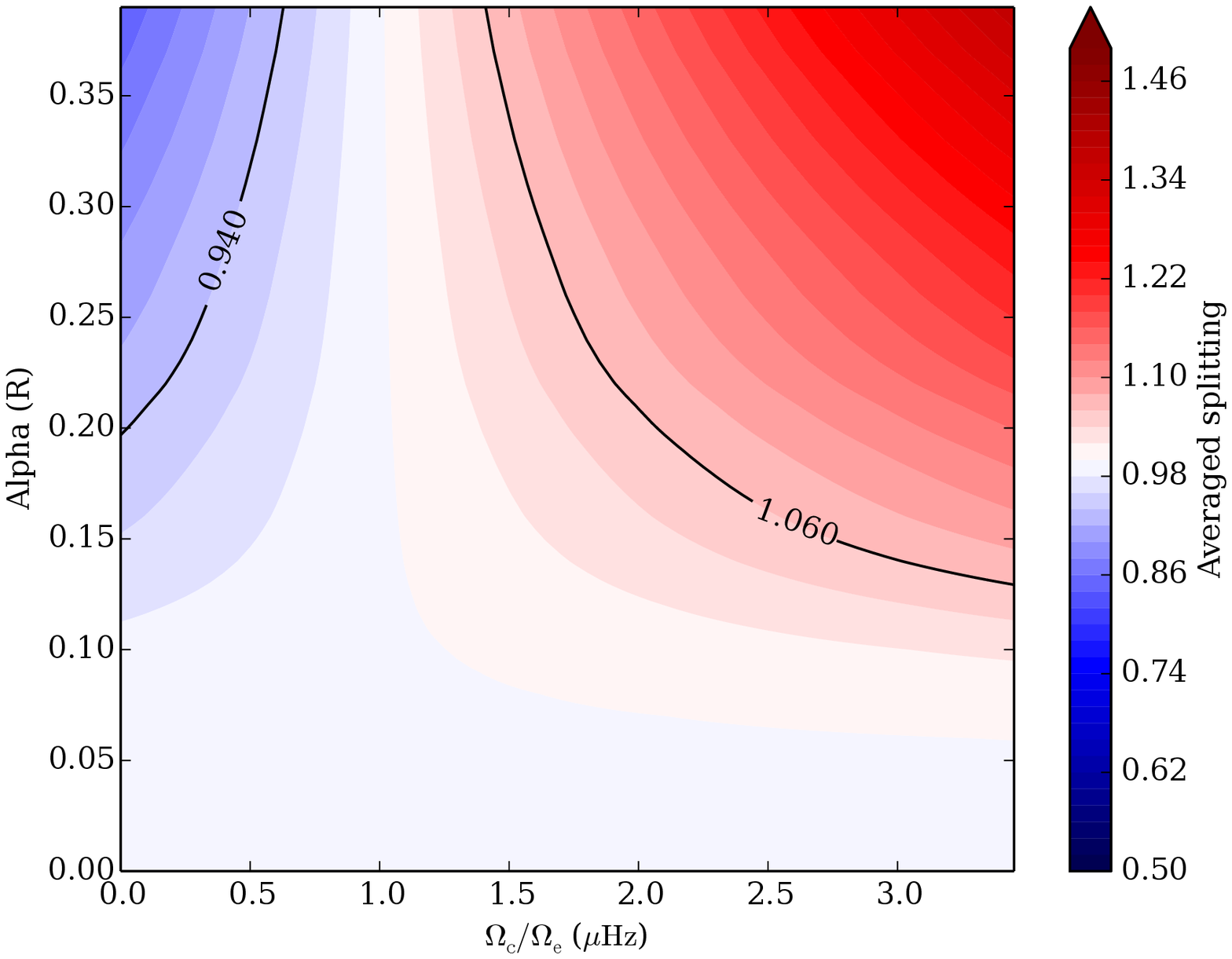} 
\includegraphics[width=88mm, clip]{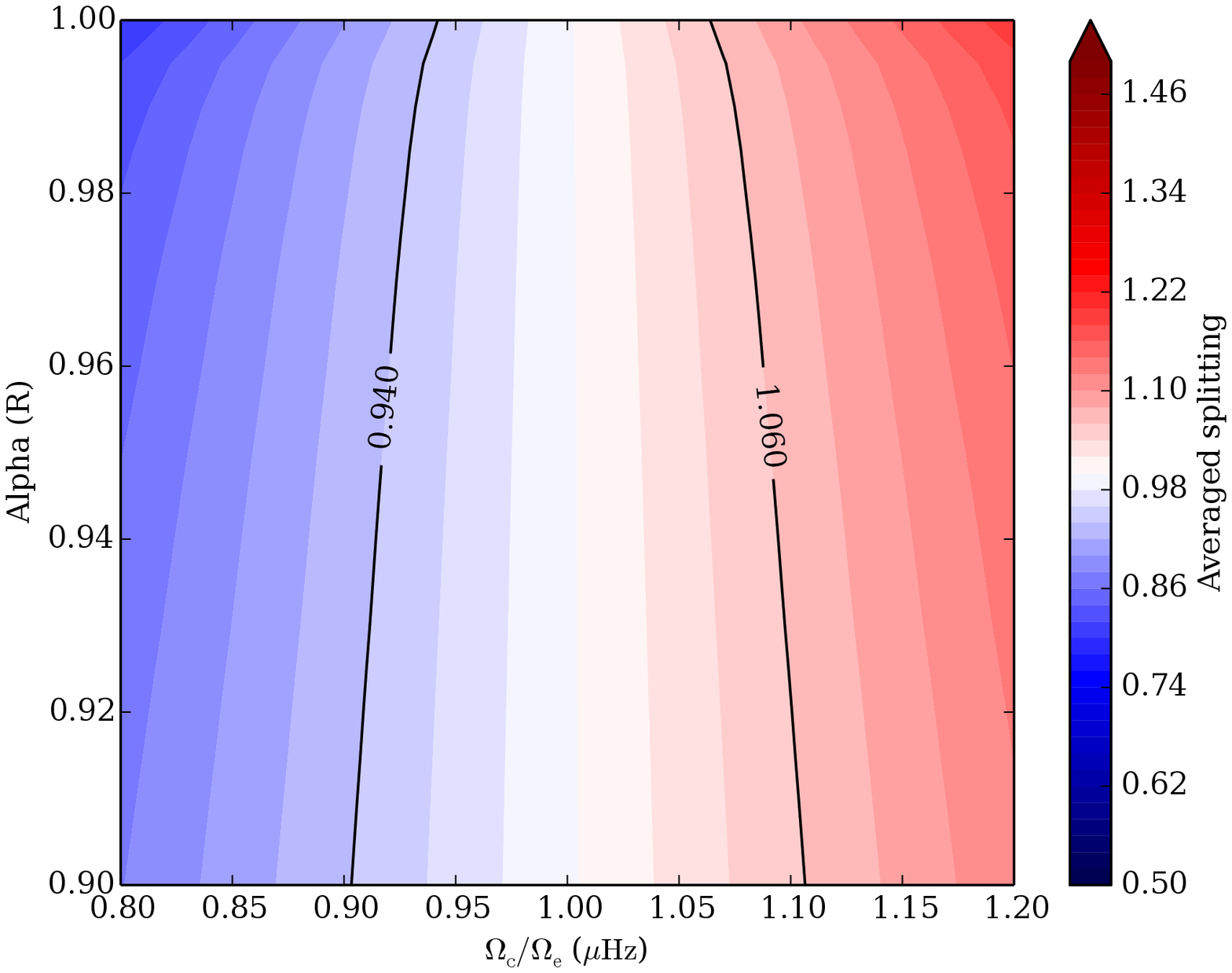} 
\caption{Averaged rotational splitting as would be measured for a star with a two level radial rotational profile with a discontiuity at a radius $\alpha$ and a central to envelope rotation rate of $\Omega_{c}/\Omega_{e}$.  The left plot corresponds to a different central rotation rate and the right plot a near surface discontinuity.  The 6\% contours that represent the precision on the 16 Cyg A measured rotational splitting are plotted in black.}
\label{fig::fastcore}
\end{figure*}
Figure \ref{fig::fastcore} also shows the equivalent of the measured averaged asteroseismic rotational splitting, as before, but for a near surface shear.  Again, regions of the parameter space that would be expected for a Sun-like star are bound by the $1\sigma$ contour.\\ 
\subsubsection{Latitudinal dependence}  
Let us now consider the latitudinal dependence in isolation.  Rotational splitting with the radial dependence removed is given as,
\begin{equation}
\delta \omega_{nlm} = m \beta_{n\ell} \int_{0}^{\pi} \mathcal{K}_{\ell m}(\theta) \Omega(\theta) \; {\rm d}\theta.
\end{equation}
We can consider the averaged latitudinal rotational kernel, $\langle \mathcal{K}_{\ell m}(\theta) \rangle$, as before by taking the weighted average of each $\mathcal{K}_{\ell m}(\theta)$ where the weight is again the reciprocal of the uncertainty in frequency.  Figure \ref{fig::surf_kern} shows the individual and averaged latitudinal rotational kernels for 16 Cyg A.  If one assumes a simple latitudinal rotational profile form of $\Omega(\theta) = A + B \cos^{2} \theta$, then for 16 Cyg A, the mean co-latitude in the measured asteroseismic rotation would be $\theta \approx 60^{\circ}$, that is around $\approx 30^{\circ}$ away from the equator.  For the range of spot latitudes observed in the Sun (near equator to ${\sim}30^{\circ}$) this mean value is entirely consistent.\\
\begin{figure}
\centering
\includegraphics[width=88mm, clip]{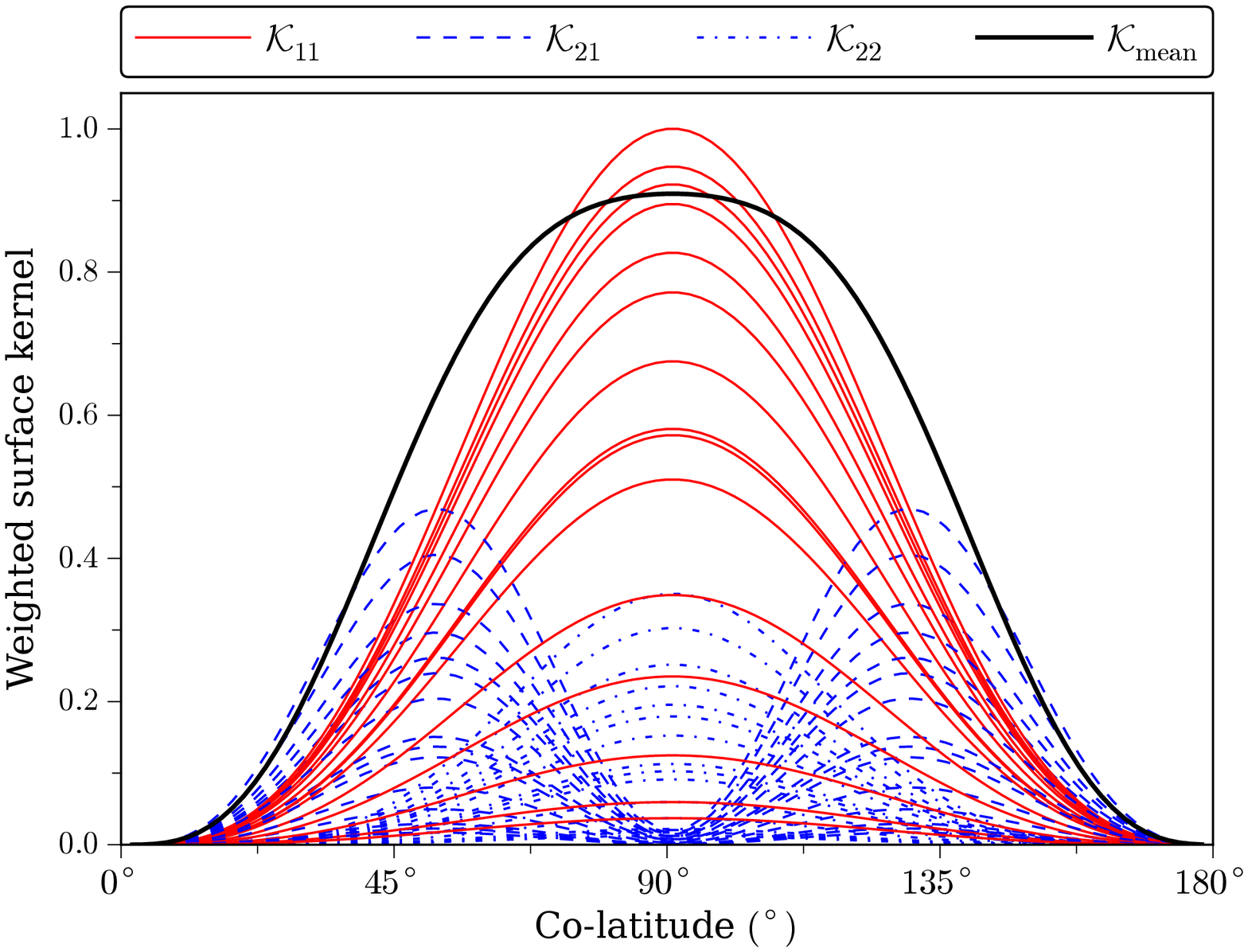} 
\caption{Co-latitudinal rotation kernels, $\mathcal{K}_{n \ell m}$, as a function of depth for 16 Cyg A.  Each kernel is weighted by the reciprocal of the uncertainty in mode frequency and normalised such that the maximum value of all the individual kernels is unity.  The strong black line shows the weighted mean of the individual kernels that represents the latitudinal rotational kernel for the averaged frequency splitting.}
\label{fig::surf_kern}
\end{figure}
\\
We conclude that for 16 Cyg A, a Sun-like star, asteroeseismic rotation is comparable to surface rotation on the assumption that there is an absence of strong radial differential rotation, in the sense of strong compared to the Sun.  This assumption is consistent with the upper limits on internal rotational gradient for move evolved stars given by \cite{2014A&A...564A..27D}.  Furthermore, surface rotation rates determined from spot modulation are sensitive to the latitude of the spot or spots that are causing the modulation.  The averaged asteroseismic rotation measurement is a reproducible measurement that is only slightly sensitive to which modes that are averaged over.  We conclude that for 16 Cyg A and B it is sensible to equate the two types of rotation.\\  
}
\subsection{Asteroseismic gyrochronology}
\begin{figure}
\includegraphics[width=88mm,clip]{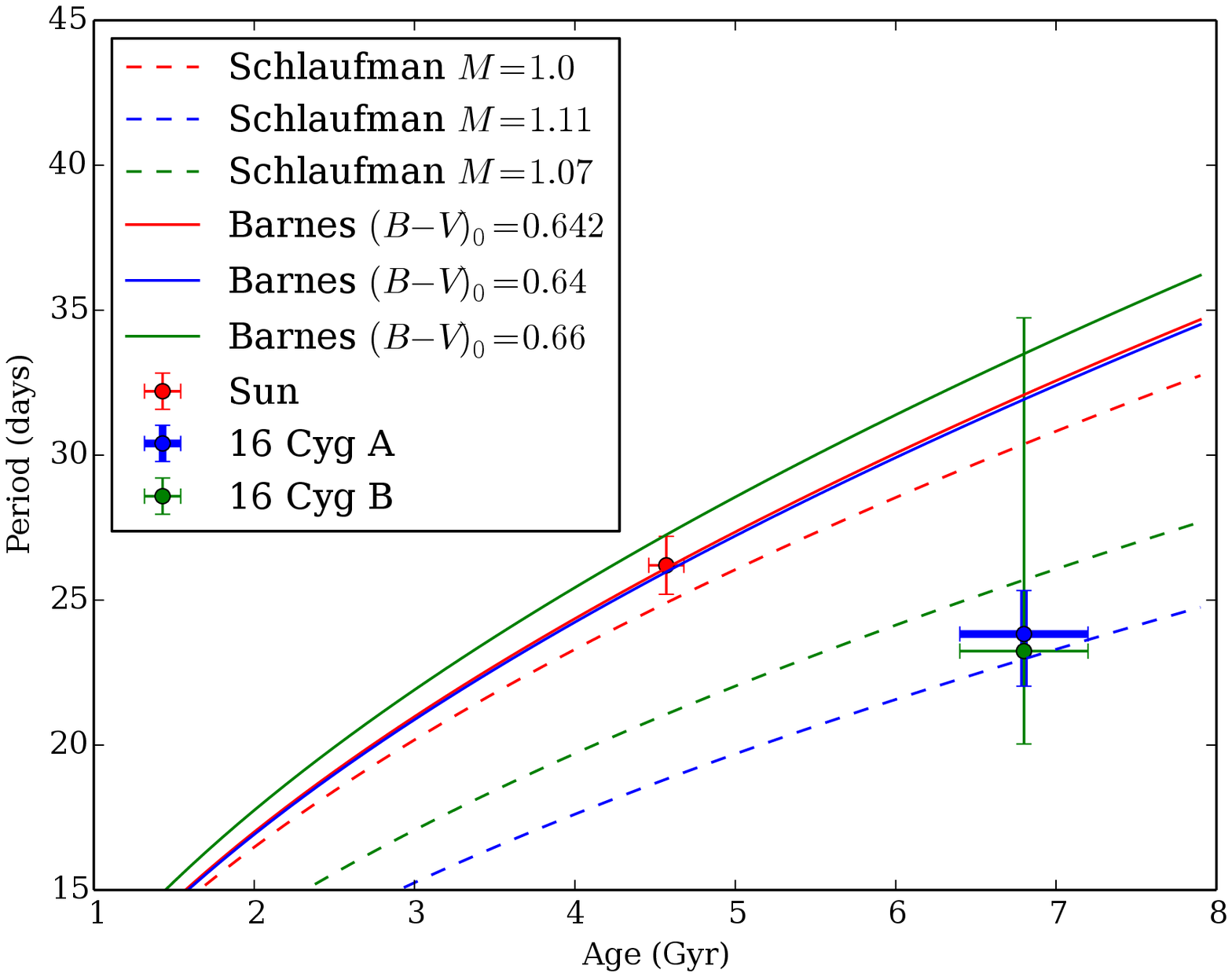}
  \caption{Asteroseismic rotation data points for 16 Cyg and the Sun on a period vs. age plot.  Predictions using two common mass-age-period relations \citep{2007ApJ...669.1167B, 2010ApJ...719..602S} are displayed.} 
  \label{fig::gyro}
\end{figure}

With the determination of the rotational periods presented here, asteroseismology has provided all three properties required to test mass-age-period relations for 16 Cyg A and B.  Here we show that the asteroseismic results give sufficient constraints so as to provide diagnostic potential.  We leave the determination of a new gyrochronology relation to future work.\\   
We selected two gyrochronology models, \cite{2007ApJ...669.1167B} and \cite{2010ApJ...719..602S}, that show deviation in the region of parameter space occupied by 16 Cyg A and B.  Figure \ref{fig::gyro} shows the models and asteroseismic data for 16 Cyg A and B and the Sun.  These gyrochronology models predict a specific relationship from mass or color, age, and period. We determined which model is preferable based on a comparison of observables to the relation specified by the model, Barnes or Schlaufman, by calculating the Bayes factor.  We then assessed the evidence for both models and created the factor as the Schlaufman evidence divided by the Barnes evidence.  The resulting Bayes factors \citep{MR0187257} are: 16 Cyg A ($\gg 1000$) ``decisive'' in favour of Schlaufman; 16 Cyg B ($ \approx 2.3$) in favour of Schlaufman but ``barely worth mentioning''.\\
The asteroseismic value for 16 Cyg A has a clear diagnostic potential and allows for an unambiguous model selection.  Given the recent work determining mass and age \todo{ \citep{2014ApJS..210....1C, 2014ApJS..214...27M} and period \citep{2014ApJS..211...24, 2014arXiv1403.7155G, 2014ApJ...790L..23D}}, in future work 16 Cyg A could provide an additional anchor when calibrating mass-age-period relations.  The ambiguity in the 16 Cyg B measurements are a result of the uncertainty in the measurement of the angle of inclination and this propagates through to the uncertainty in the rotational period.\\
Given the Schlaufmann gyrochronology relation we can define a prior probability distribution for the period of 16 Cyg B in order to better constrain the angle of inclination.  We define a uniform prior between the $3\sigma$ limits of the Schlaufmann relation, that is a uniform prior between 21 and 30 days.  Repeating the analysis above with the addition of this prior gives a consistent but much better constrained result.  We now have the angle of inclination for 16 Cyg B of $36_{-4}^{+5}$ degrees.\\
\subsection{Inclinations and orbital properties}
Systems such as 16 Cyg constitute a promising laboratory to probe star-planet interactions such as tides.  Because of the large separation between 16 Cyg A and 16 Cyg B, of around 750 AU \citep{2013AJ....146..108P}, and the strength of tidal interactions falling with distance cubed, we can consider the tidal effects in the star-planet system 16 Cyg B-Bb as if it is isolated from 16 Cyg A. This reduces the problem to the standard case of a binary system in tidal interaction. We can then apply the method described in detail in \cite{Hut1981} and \cite{Mardling2011} to predict the circularisation time for the orbit and the alignment and synchronisation timescales for 16 Cyg B and Bb.\\
First, we adopt the asteroseismic mass and radius for 16 Cyg B \citep[][Table 1]{2012ApJ...748L..10M}. Next, following \cite{2013AJ....146..108P}, we assume that $M_{\rm p}=2.38\,M_{\rm J}$, where $M_{\rm p}$ and $M_{\rm J}$ are respectively the mass of 16 Cyg Bb and of Jupiter.  This leads us to its approximate radius, that $R_{\rm p}\approx R_{\rm J}$ using the mass-radius relation for giant planets \citep{CLB2011}. Next, assuming the usual values for tidal dissipation in low-mass stars and giant planets \citep{BO2009}, we are able to predict the time scales of circularisation for the orbit, the alignment, and synchronisation times for 16 Cyg B, and 16 Cyg Bb given in Table 3.\\
These results suggest that the rotation of the planet 16 Cyg Bb is synchronised with its orbital motion (tidally locked) while its spin is still evolving to become aligned with the total angular momentum of the system, assuming it had 
an initial non zero obliquity.\\
The timescale of star-planet orbit circularisation due to tidal effects is much greater than the age of the system, over 6 orders of magnitude greater, which is consistent with the observed high level of eccentricity.\\
For the purpose of the tidal effects we considered the B-Bb system as isolated but it is still possible that other dynamical effects can impact the state of the system.  It is unclear whether the eccentricity of the star-planet orbit is due to ongoing interactions with 16 Cyg A \citep{1999PASP..111..321H, 1997Natur.386..254H,1997ApJ...477L.103M} or primordial \citep{2012ApJ...757...18A}.  Our own dynamical orbital calculations suggest that about 5 to 10\% of the orbital phase space consistent with the constraints in Table 1 result in dynamical interactions between 16 Cyg B-Bb and 16 Cyg A (Kozai cycling) that produce significant ($e > 0.5$) eccentricity on a timescale of a Gyr.\\
The stellar orbit of B around A is given as inclined at 100 to 160 degrees \citep{1999PASP..111..321H}, which is consistent, but does not discriminate, with the other angles measured here.  In addition, the angle of inclination for B of $36^{+5}_{-4}$ degrees and the Bb orbit angle of inclination of $45\pm 1$ degrees gives a low projected obliquity when assuming co-rotation. This low projected obliquity corresponds to a true obliquity mode of 16 degrees, with a 68\% upper limit of 39 degrees.  This obliquity may be simply primordial or may be a result of the orbit's phase in a Kozai cycle due to the gravitational interaction with 16 Cyg A, with other phases having higher obliquity; in either case, the tidal alignment timescale is much larger than the age of the system.
\begin{table}
 \caption{Tidal dynamical time scales of the 16 Cyg B-Bb system}
  \label{tab::dyn}
 \begin{center}
 \begin{tabular}{ccc}
 \hline
 & 16 Cyg B & 16 Cyg Bb \\
 \hline\hline
 Circularisation time scale (Gyr) & $3.455\,\times 10^{7}$ & $3.455\, \times 10^{7}$ \\
 Alignment time scale (Gyr) & $2.913\, \times 10^{5}$ & $7.131$ \\
 Synchronisation time scale (Gyr) & $1.458\, \times 10^{5}$ & $3.566$ \\
 \hline \\
 \end{tabular}
 \end{center}
\end{table}

\section{Conclusions}
\guy{We have tried to measure surface rotation for the two stars 16 Cyg A and 16 Cyg B.  In the absence of surface rotation we have used the established technique of asteroseismology to determine the rotational properties.  We have measured the magnitude of the projected rotational splittings in frequency ($0.411 \pm 0.013 \rm \, \mu Hz$, $0.274 \pm 0.017 \rm \, \mu Hz$) and the angles of inclination ($56^{+6}_{-5} \, ^{\circ}$, $36^{+17}_{-7} \, ^{\circ}$).  This has allowed us to determine the asteroseismic $v \sin i$ ($2.23 \pm 0.07 \rm \, km \, s^{1}$, $1.35 \pm 0.08 \rm \, km \, s^{-1}$) and the period of rotation of ($23.8^{+1.5}_{-1.8} \rm \, days$, $23.2^{+11.5}_{-3.2} \rm \, days$). We have checked our result by applying our method to solar data from the red channel of the SPM/VIRGO instrument and find good agreement with the known solar rotation rate.  We have also discussed the consequences of comparing asteroseismic and surface rotation periods for Sun-like stars.\\}
We have used the results on rotation to evaluate two gyrochronology relations for the 16 Cyg system\guy{, using an adopted age of $6.8 \pm 0.4 \rm \, Gyr$}.  While the results for 16 Cyg B provide little diagnostic potential, the parameters of 16 Cyg A are sufficiently well known that we are able to tension gyrochronology relations.  We find a decisive result for the Schlaufman relation in a Bayesian model comparison.  \guy{In future work, under the assumption of solar-like rotation, we suggest that 16 Cyg A could be used as an anchor for calibrating gyrochronology relations.  \todo{We note that the Schlaufman relation gives ages of 7.3 Gyr and 5.5 Gyr for the A and B components, respectively, when using the asteroseismic masses as input.}}\\
We have discussed the planetary system dynamics with the additional information provided by the rotation results.  We have calculated the tidal dynamical time scales for the eccentric 16 Cyg B-Bb system, finding that the orbit circularisation time scale is much greater than the age of the system.  Further more, we have calculated the projected and true obliquities of the planetary system and find results that are consistent with a low true obliquity.  This low obliquity is required in a dynamical system that is dominated by Kozai cycling.  The current measured state of the system is then consistent with a high eccentricity, low obliquity orbit where 16 Cyg A is driving the Kozai cycling.

\section{Acknowledgements}
The authors thank A. Jim\'{e}nez for providing the SoHO/VIRGO/SPM time series. SoHO is a space mission of international cooperation between ESA and NASA.  GRD, RAG, TC, ans SM received funding from the CNES GOLF and CoRoT grants at CEA. RAG also acknowledges the ANR (Agence Nationale de la Recherche, France) program IDEE (n° ANR-12-BS05-0008) ``Interaction Des \'Etoiles et des Exoplan\`etes '' .  The research leading to these results has received funding from the European Community’s Seventh Framework Programme ([FP7/2007-2013]) under grant agreement no. 312844 (SPACEINN) and under grant agreement no. 269194 (IRSES/ASK).  GRD, WJC, TLC, YE, and RH acknowledge the support of the UK Science and Technology Facilities Council (STFC).  DS is supported by the Australian Research Council.  SB acknowledges partial support from NSF grant AST-1105930 and NASA grant NNX13AE70G.  Funding for the Stellar Astrophysics Centre is provided by The Danish National Research Foundation (Grant agreement no.: DNRF106). The research is supported by the ASTERISK project (ASTERoseismic Investigations with SONG and Kepler) funded by the European Research Council (Grant agreement no.: 267864).  This work was supported partly by the Programme National de Plan\'{e}tologie (CNRS/INSU) and the Campus Spatial de l'Universit\'{e} Paris Diderot.  
\bibliographystyle{mn2e_new}
\bibliography{refs}

\appendix
\section{Modes treated in the peakbagging process}
\begin{table}
 \caption{Mode frequencies for 16 Cyg A with 68\% credible regions}
  \label{tab::freqa}
 \begin{center}
 \begin{tabular}{cccc}
 \hline
 $n$ & $\ell$ & frequency & 68\% credible region  \\
 &  & $\mu \rm \; Hz$ & $\mu \rm \; Hz$ \\
 \hline\hline
12 & 2 & 1488.24 & 0.51 \\
13 & 0 & 1495.00 & 0.07 \\
13 & 1 & 1541.92 & 0.07 \\
13 & 2 & 1591.29 & 0.19 \\
14 & 0 & 1598.69 & 0.07 \\
14 & 1 & 1645.06 & 0.09 \\
14 & 2 & 1694.17 & 0.17 \\
15 & 0 & 1700.91 & 0.08 \\
15 & 1 & 1747.15 & 0.08 \\
15 & 2 & 1795.75 & 0.11 \\
16 & 0 & 1802.31 & 0.07 \\
15 & 3 & 1838.52 & 0.67 \\
16 & 1 & 1848.98 & 0.05 \\
16 & 2 & 1898.26 & 0.10 \\
17 & 0 & 1904.61 & 0.06 \\
16 & 3 & 1941.22 & 0.56 \\
17 & 1 & 1952.00 & 0.05 \\
17 & 2 & 2001.67 & 0.08 \\
18 & 0 & 2007.58 & 0.05 \\
17 & 3 & 2045.98 & 0.37 \\
18 & 1 & 2055.52 & 0.05 \\
18 & 2 & 2105.31 & 0.06 \\
19 & 0 & 2110.91 & 0.04 \\
18 & 3 & 2149.94 & 0.13 \\
19 & 1 & 2159.15 & 0.04 \\
19 & 2 & 2208.90 & 0.06 \\
20 & 0 & 2214.22 & 0.05 \\
19 & 3 & 2253.53 & 0.16 \\
20 & 1 & 2262.54 & 0.05 \\
20 & 2 & 2312.54 & 0.09 \\
21 & 0 & 2317.32 & 0.05 \\
20 & 3 & 2357.39 & 0.19 \\
21 & 1 & 2366.25 & 0.06 \\
21 & 2 & 2416.25 & 0.13 \\
22 & 0 & 2420.90 & 0.08 \\
21 & 3 & 2462.08 & 0.38 \\
22 & 1 & 2470.30 & 0.08 \\
22 & 2 & 2520.46 & 0.21 \\
23 & 0 & 2525.07 & 0.16 \\
22 & 3 & 2566.97 & 0.61 \\
23 & 1 & 2574.78 & 0.13 \\
23 & 2 & 2624.32 & 0.32 \\
24 & 0 & 2629.20 & 0.18 \\
23 & 3 & 2669.76 & 1.04 \\
24 & 1 & 2679.87 & 0.19 \\
24 & 2 & 2730.23 & 0.89 \\
25 & 0 & 2733.61 & 0.46 \\
25 & 1 & 2784.22 & 0.35 \\
25 & 2 & 2835.34 & 1.15 \\
26 & 0 & 2838.40 & 0.78 \\
26 & 1 & 2891.27 & 0.74 \\
26 & 2 & 2941.48 & 1.54 \\
27 & 0 & 2945.32 & 1.18 \\
27 & 1 & 2996.38 & 1.19 \\
 \hline \\
 \end{tabular}
 \end{center}
\end{table}
\begin{table}
 \caption{Mode frequencies for 16 Cyg B with 68\% credible regions}
  \label{tab::freqb}
 \begin{center}
 \begin{tabular}{cccc}
 \hline
 $n$ & $\ell$ & frequency & 68\% credible region  \\
 &  & $\mu \rm \; Hz$ & $\mu \rm \; Hz$ \\
 \hline\hline
12 & 2 & 1686.42 & 0.31 \\
13 & 0 & 1695.07 & 0.09 \\
13 & 1 & 1749.21 & 0.1 \\
13 & 2 & 1804.17 & 0.27 \\
14 & 0 & 1812.43 & 0.1 \\
14 & 1 & 1866.52 & 0.12 \\
14 & 2 & 1921.21 & 0.16 \\
15 & 0 & 1928.90 & 0.07 \\
14 & 3 & 1970.96 & 5.14 \\
15 & 1 & 1982.59 & 0.07 \\
15 & 2 & 2036.67 & 0.14 \\
16 & 0 & 2044.28 & 0.06 \\
15 & 3 & 2085.37 & 1.5 \\
16 & 1 & 2098.08 & 0.06 \\
16 & 2 & 2152.42 & 0.1 \\
17 & 0 & 2159.58 & 0.06 \\
16 & 3 & 2200.58 & 1.22 \\
17 & 1 & 2214.17 & 0.06 \\
17 & 2 & 2268.96 & 0.08 \\
18 & 0 & 2275.95 & 0.05 \\
17 & 3 & 2319.12 & 0.37 \\
18 & 1 & 2331.14 & 0.04 \\
18 & 2 & 2386.26 & 0.06 \\
19 & 0 & 2392.71 & 0.04 \\
18 & 3 & 2436.66 & 0.3 \\
19 & 1 & 2448.25 & 0.04 \\
19 & 2 & 2503.50 & 0.06 \\
20 & 0 & 2509.67 & 0.04 \\
19 & 3 & 2554.15 & 0.15 \\
20 & 1 & 2565.40 & 0.04 \\
20 & 2 & 2620.56 & 0.07 \\
21 & 0 & 2626.40 & 0.04 \\
20 & 3 & 2671.72 & 0.17 \\
21 & 1 & 2682.40 & 0.05 \\
21 & 2 & 2737.74 & 0.08 \\
22 & 0 & 2743.33 & 0.06 \\
21 & 3 & 2789.15 & 0.28 \\
22 & 1 & 2799.73 & 0.06 \\
22 & 2 & 2855.63 & 0.12 \\
23 & 0 & 2860.72 & 0.1 \\
22 & 3 & 2906.87 & 0.44 \\
23 & 1 & 2917.79 & 0.1 \\
23 & 2 & 2973.56 & 0.24 \\
24 & 0 & 2978.50 & 0.15 \\
23 & 3 & 3025.06 & 1.13 \\
24 & 1 & 3036.06 & 0.15 \\
24 & 2 & 3093.04 & 0.51 \\
25 & 0 & 3096.85 & 0.42 \\
24 & 3 & 3144.04 & 1.42 \\
25 & 1 & 3154.31 & 0.29 \\
25 & 2 & 3213.40 & 1.54 \\
26 & 0 & 3214.93 & 1.04 \\
26 & 1 & 3273.17 & 0.64 \\
26 & 2 & 3333.06 & 2.74 \\
27 & 0 & 3334.22 & 1.9 \\
27 & 1 & 3393.45 & 0.77 \\
 \hline \\
 \end{tabular}
 \end{center}
\end{table}

\label{lastpage}
\end{document}